\documentclass[preprint]{JHEP}
\usepackage{epsf}
\def\eq#1{{eq.~(\ref{#1})}}
\def\eqs#1#2{{eqs.~(\ref{#1})--(\ref{#2})}}

\let\vev\VEV
\def\abs#1{\left| #1\right|}
\def\mod#1{\abs{#1}}
\def\Im{\mathop{\mbox{Im}}}
\def\Re{\mathop{\mbox{Re}}}
\def\Tr{\mathop{\mbox{Tr}}\,}
\def\etal{{\it et al.}}

\def\qq{$\vev{\bar qq}$~}
\def\GG{$\vev{\alpha_s GG/\pi}$~}

\def\ee{$\varepsilon'/\varepsilon$~}
\newcommand{\bea}{\begin{eqnarray}}
\newcommand{\beq}{\begin{equation}}
\newcommand{\eea}{\end{eqnarray}}
\newcommand{\eeq}{\end{equation}}
\newcommand{\nnu}{\nonumber}

\title{$\varepsilon '/\varepsilon$ at $O (p^4)$
 in the Chiral Expansion} 
\author{S. Bertolini$^{\dag \S}$, J.O. Eeg$^{\ddag}$,
M. Fabbrichesi$^{\dag \S}$ and E.I. Lashin$^{\dag}$\thanks{Permanent address:
Ain Shams University, Faculty of Science, Dept. of Physics, Cairo, Egypt.}\\
$^{\dag}$ INFN, Sezione di Trieste\\
$^{\S}$ Scuola Internazionale Superiore di Studi Avanzati\\
via Beirut 4, I-34013 Trieste, Italy.\\
$^{\ddag}$ Fysisk Institutt, Universitetet i Oslo\\
N-0316 Oslo, Norway.}

\abstract{After updating the determination of the combination of
Koba\-ya\-shi-Mas\-kawa elements $ \Im V_{td}V^*_{ts}$ according to
our new estimate of the parameter $\widehat B_K$,
we study the $CP$-violating ratio $\varepsilon'/\varepsilon$
by means of hadronic matrix elements computed to $O(p^4)$ in the
chiral expansion. It is the first time
that this order in chiral perturbation theory is included.
We also discuss the relevance of some $O(p^2)$ terms 
that are generally neglected
in the calculation of the electroweak penguin matrix elements. 
The  most important effect of this improved analysis
is the substantial reduction (20\%) of the leading  
electroweak penguin matrix element 
and accordingly a reduced cancelation between the electroweak
and gluon penguin contributions. 
The ratio $\varepsilon'/\varepsilon$ is thus larger than previously estimated
and its predicted value enjoys a smaller uncertainty. We find 
$\varepsilon'/\varepsilon = 1.7\ ^{+1.4}_{-1.0} \times 10^{-3}$.
Values positive and of the order of $10^{-3}$ are therefore
preferred.}
 
\preprint{SISSA 20/97/EP \\
May 1997 \\
Revised, September 1997}

\begin{document}

\section {Introduction}	

The real part of the ratio \ee is a fundamental
parameter of the standard model that measures the amount of 
direct $CP$ violation in the decay of a neutral kaon in two 
 pions.

Most recent
estimates (see refs.~\cite{munich,rome,III})
 have so far agreed that values of the order of $10^{-4}$
are to be preferred within the standard model. The reason for
the smallness of these values  resides in the cancelation between the
contributions of two classes of diagrams: the gluon and the electroweak
penguins. Moreover, because 
each class gives a contribution of order  $10^{-3}$
and of the opposite sign, the resulting value of order  $10^{-4}$ is
plagued by an uncertainty of at least 100\%---as it can be readily understood
by assigning an optimistic uncertainty of 10\% at each of the two
contributions.

Our previous estimate of \ee~\cite{III}, was based on the  
leading order (LO) lagrangian containing
$O(p^0)$  and $O(p^2)$ terms, where the corresponding
coefficients  were calculated by means of the
chiral quark model ($\chi$QM)~\cite{QM,BBdeR}.
The $O(p^4)$ mesonic one-loop corrections 
for the $K^0\to\pi\pi$ matrix elements were then calculated
via the LO chiral lagrangian.
Finally, the fit of the $\Delta I =1/2$ selection rule in 
$K^0\to\pi\pi$ decays allowed us to 
constrain the ranges of the free parameters involved.
Nonetheless,
in ref.~\cite{III} we were  forced to
assign a rather large uncertainty to our prediction of \ee 
because of the almost complete cancelation 
between gluon and electroweak
contributions. 

We now take our analysis one step
further by including, in the
calculation of the hadronic matrix elements,
the terms proportional
to the quark current masses and up to  $O(p^4)$ in momenta. 
These are terms next-to-leading
order (NLO) in the  $\Delta S = 1$ chiral lagrangian determining \ee. 

As we shall discuss, the inclusion of the
terms proportional to the current quark masses and those of $O(p^4)$ 
shows that the cancelation we find in the LO
is substantially reduced.
As a result, the preferred central value for 
\ee is now above $10^{-3}$, even though values 
of \ee of order $10^{-4}$ cannot be excluded.

In summary, the new elements that we have introduced
in the present analysis are:
\begin{itemize}
\item
a term proportional to the quark current masses $m_q$, 
to be added to the $O(p^2)$ corrections to 
the leading $O(p^0)$ part of the lagrangian
which characterizes the electroweak penguin contribution;
\item
the  $O(p^4)$ matrix elements of the relevant 
$\Delta S = 1$ four-quark operators 
directly computed in the $\chi$QM via quark integration
(these matrix elements correspond to
linear combinations of chiral coefficients of the
 $O(p^4)$ chiral lagrangian). This order has never been
included before;
\item
the contribution of the chromomagnetic 
operator $Q_{11}$ (which arise at $O(p^4)$), 
first discussed in ref. \cite{BEF}
but so far never systematically included;
\item
updated values for $\alpha_s(m_Z)$ and $m_t$;
\item
updated ranges of the $\chi$QM parameters \qq and \GG  obtained
from our recent $O(p^4)$ fit of the $\Delta I = 1/2$ selection
rules in kaon decays~\cite{V}.
\end{itemize}

The most important consequences of these
improvements are summarized by the following results:
\begin{itemize}
\item 
the effect of consistently
including all $O(p^2)$ corrections to the $O(p^0)$ term
in the LO chiral lagrangian reduces by about 20\% the
contribution of the electroweak penguin operators thus making 
the central value of \ee larger; 
\item
the uncertainty of the analysis is reduced as a consequence of
the less effective cancelation between the 
gluon and electroweak penguin contributions; 
\item 
the NLO
$O(p^4)$ corrections are well under control and amount to a 10\% effect
on the predicted \ee.
\end{itemize}

Such an encouraging outcome makes us look forward to
the  next generation of experiments, under way at
CERN, FNAL and DA$\Phi$NE, that will
reach the sensitiveness of $(1\div2) \times 10^{-4}$
and hopefully
 resolve the present uncertainty of the results of CERN (NA31)~\cite{CERN}
 \beq
 \mbox{Re} \: \left( \varepsilon'/\varepsilon \right) = 
 (23 \pm 7) \times 10^{-4}
 \eeq
 and FNAL (E731)~\cite{SLAC}
 \beq
 \mbox{Re} \: \left( \varepsilon'/\varepsilon \right) = 
 (7.4 \pm 6.0) \times 10^{-4} \, .
 \eeq

In the following subsections, before going into the detailed analysis, we
summarize the notation and the framework
used in our previous work.

\subsection{Basic Formulae}

 The $\Delta S=1$
quark effective lagrangian at a scale $\mu < m_c$ can be written
as~\cite{GW}
 \beq
{\cal L}_{\Delta S = 1} = -
\frac{G_F}{\sqrt{2}} V_{ud}\,V^*_{us} \sum_i \Bigl[
z_i(\mu) + \tau y_i(\mu) \Bigr] Q_i (\mu) 
 \, , \label{Lquark}
\eeq
where $i=1,...,10$.
The functions $z_i(\mu)$ and $y_i(\mu)$ are the
 Wilson coefficients (known to the NLO in the strong coupling 
expansion~\cite{NLO1}) and $V_{ij}$ the
Koba\-ya\-shi-Mas\-kawa (KM) matrix elements; 
$\tau = - V_{td}V_{ts}^{*}/V_{ud}V_{us}^{*}$.

The $Q_i$ are the effective
four-quark operators obtained by integrating out in the standard
model the vector bosons and the heavy quarks $t,\,b$ and $c$. They are 
by now the standard basis for the description of $\Delta S=1$
transitions and can be written as
\beq
\begin{array}{rcl}
Q_{1} & = & \left( \overline{s}_{\alpha} u_{\beta}  \right)_{\rm V-A}
            \left( \overline{u}_{\beta}  d_{\alpha} \right)_{\rm V-A}
\, , \\[1ex]
Q_{2} & = & \left( \overline{s} u \right)_{\rm V-A}
            \left( \overline{u} d \right)_{\rm V-A}
\, , \\[1ex]
Q_{3,5} & = & \left( \overline{s} d \right)_{\rm V-A}
   \sum_{q} \left( \overline{q} q \right)_{\rm V\mp A}
\, , \\[1ex]
Q_{4,6} & = & \left( \overline{s}_{\alpha} d_{\beta}  \right)_{\rm V-A}
   \sum_{q} ( \overline{q}_{\beta}  q_{\alpha} )_{\rm V\mp A}
\, , \\[1ex]
Q_{7,9} & = & \frac{3}{2} \left( \overline{s} d \right)_{\rm V-A}
         \sum_{q} \hat{e}_q \left( \overline{q} q \right)_{\rm V\pm A}
\, , \\[1ex]
Q_{8,10} & = & \frac{3}{2} \left( \overline{s}_{\alpha} 
                                                 d_{\beta} \right)_{\rm V-A}
     \sum_{q} \hat{e}_q ( \overline{q}_{\beta}  q_{\alpha})_{\rm V\pm A}
\, , 
\end{array}  
\label{Q1-10} 
\eeq
where $\alpha$, $\beta$ denote color indices ($\alpha,\beta
=1,\ldots,N_c$) and $\hat{e}_q$  are quark charges. Color
indices for the color singlet operators are omitted. 
The subscripts $(V\pm A)$ refer to
$\gamma_{\mu} (1 \pm \gamma_5)$ in the quark currents.

Starting from the quark effective operators in \eq{Q1-10} and based
on the $\chi$QM approach we have
given in ref. \cite{I} a discussion of their 
bosonic representation and the determination of the corresponding
$O(p^2)$ chiral coefficients. 
In the present paper we complete the $O(p^2)$ part of the analysis
by including a term proportional to the current quark masses
previously neglected.

Two additional operators enter when considering $O(p^4)$ matrix elements:
\bea
Q_{11} & = & \frac{g_s}{16 \pi^2} \: \bar s \: \left[ 
m_d \left(1 + \gamma_5 \right) + m_s \left(1 - \gamma_5 \right) \right] 
\: \sigma \cdot G \: d
\, , 
\label{Q11}\\[1ex]
Q_{12} & = & \frac{e}{16 \pi^2} \: \bar s \: \left[ 
m_d \left(1 + \gamma_5 \right) + m_s \left(1 - \gamma_5 \right) \right] 
\: \sigma \cdot F \: d
\, .
\label{Q12}
\eea
The bosonization of the chromomagnetic operator $Q_{11}$ is discussed
in ref.~\cite{BEF} where it is shown that the first non-vanishing
contribution arises at $O(p^4)$ and it is found to be further
suppressed by a factor $m_\pi^2/m_K^2$ compared to the naive expectation.
Our complete NLO analysis shows the subleading role 
of these operators for $K^0\to\pi\pi$.
 
For completeness we mention the so called self-penguin contribution to
\ee considered in ref.~\cite{BEg}. This contribution has not been included in
any systematic NLO short-distance analysis. Since the effect
on \ee is of order $10^{-4}$, and thereby well within
the errors quoted, it is omitted altogether.

The quantity $\varepsilon'/\varepsilon$ can be written as
\beq
\frac{\varepsilon'}{\varepsilon} =  
e^{i (\delta_2 - \delta_0 + \pi/4)}
\frac{G_F \omega}{2\mod{\epsilon}\Re{A_0}} \:
\mbox{Im}\, \lambda_t \: \:
 \left[ \Pi_0 - \frac{1}{\omega} \: \Pi_2 \right] \, ,
\label{eps'}
 \eeq
where, referring to the $\Delta S=1$ quark lagrangian of \eq{Lquark},
\bea
 \Pi_0 & = & \frac{1}{\cos\delta_0} \sum_i y_i \, 
\Re\langle  Q_i  \rangle _0\ (1 - \Omega_{\eta +\eta'}) 
\label{PI0}
\\
 \Pi_2 & = & \frac{1}{\cos\delta_2} \sum_i y_i \, 
\Re\langle Q_i \rangle_2 \ ,
\label{PI2}
\eea
and
\beq
\mbox{Im}\, \lambda_t \equiv \Im V_{td}V^*_{ts} \, .
\eeq
Since Im $ \lambda_u =0$ according to the standard conventions,
the short-distance component of $\varepsilon'/\varepsilon$
is determined by the Wilson coefficients $y_i$. 
Following the approach of
ref.~\cite{NLO1}, $y_1(\mu)=$ $y_2(\mu)=0$. 
As a consequence, the matrix elements of $Q_{1,2}$ do not directly
enter the determination of $\varepsilon'/\varepsilon$.

Experimentally the phases $\delta_I$ are obtained in terms
of the $\pi$-$\pi$ S-wave scattering lenght~\cite{phase02exp}
at the $m_K$ scale. The values so derived give to a few degrees uncertainty
$\delta_0 \simeq 37^0$ and $\delta_2 \simeq - 8^0$,
thus obtaining with good accuracy
\bea
\cos\delta_0 &\simeq& 0.8 \nnu\\
\cos\delta_2 &\simeq& 1.0\ .
\label{cosdelta02exp}
\eea
As a consequence the $I=0$ amplitude includes a 20\% enhancement
from the rescattering phase.
In addition, one has
$\delta_0 - \delta_2 - \pi/4 \simeq 0$ thus making the 
$\varepsilon'/\varepsilon$ ratio approximately real.

We take as input values
\beq
\frac{G_F \omega}{2\mod{\epsilon}\Re{A_0}} \simeq 349 \ \mbox{GeV}^{-3},
\qquad 
\omega = 1/22.2\ , \qquad \Omega_{\eta + \eta'} = 0.25 \pm 0.05 \ .
\label{values}
\eeq

The large value in \eq{values} for $1/\omega$ comes from
the $\Delta I =1/2$ selection rule. In refs. \cite{V,II} we have 
shown that such a rule is
well reproduced by the chiral quark model ($\chi$QM)
evaluation of the hadronic matrix elements. 

The quantity $\Omega_{\eta +\eta'}$ 
includes the effect on $\Pi_2$ of the isospin-breaking 
mixing between $\pi^0$ and the etas (see refs.~\cite{BG} and \cite{DL}).
Being this effect proportional to $\Pi_0$, it is
written for notational convenience as a part of the $\Pi_0$ contribution. 
Its uncertainty affects the error 
in the final estimate of \ee by about 10\%. 

As the precise values of $\Re A_0$ and $\omega$  
depend on the  choice of the $\chi$QM parameters---the quark and 
gluon condensates and the constituent quark mass $M$---we have
used in the $\varepsilon'/\varepsilon$ estimate
the range of values that reproduce at $O(p^4)$ the
$\Delta I = 1/2$ rule with a 20\% accuracy~\cite{V}.

\subsection{The Chiral Quark Model}

Effective quark models of QCD can be derived in the framework of
the extended Nambu-Jona-Lasinio (ENJL) model of chiral symmetry
breaking~\cite{BBdeR}.
Among them we have utilized the $\chi$QM~\cite{QM}, 
which is the mean-field approximation of the full ENJL and 
in which an effective interaction term 
\beq
 {\cal{L}}^{\rm int}_{\chi \mbox{\scriptsize QM}} = 
- M \left( \overline{q}_R \; \Sigma q_L +
\overline{q}_L \; \Sigma^{\dagger} q_R \right) \, ,
\label{M-lag}
\eeq
is added to a QCD lagrangian 
whose propagating fields are the $u,d,s$ quarks
in the fixed background of soft gluons.

In the factorization approximation, matrix elements of the four
 quark operators are written in terms of better known
quantities like quark currents and densities. As a typical example, a matrix
element of $Q_6$ can be written as
\bea
\langle \pi^+ \pi^-|Q_6| K^0 \rangle & = & 
 2 \ \langle \pi^-|\overline{u}\gamma_5 d|0 \rangle
\langle \pi^+|\overline{s} u |K^0 \rangle 
- 2 \ \langle \pi^+ \pi^-|\overline{d} d|0 \rangle  \langle 0|\overline{s} 
\gamma_5 d |K^0 \rangle
\nnu \\
& & +\ 2 \ \left[ \langle 0|\overline{s} s|0 \rangle \, - \, 
\langle 0|\overline{d}d|0 \rangle
\right] \,   \langle \pi^+ \pi^-|\overline{s}\gamma_5 d |K^0 \rangle \, .
\label{MQ6}
\eea

The matrix elements (building blocks) like 
$\langle0|\,\overline{s} \gamma_5 u\, |K^+(k)\rangle$,
$\langle \pi^+(p_+)|\,\overline{s} d\, |K^+(k)\rangle$ and  
$\langle 0|\,\overline{s} \gamma^\mu \left(1 - \gamma_5\right) u\,|K^+(k)
\rangle$ are then evaluated to $O (p^4)$ within the model and
the four-quark $K \rightarrow \pi \pi$ matrix elements
are obtained by assembling the building blocks at the given order
in momenta.

Non-perturbative gluonic corrections, are included when calculating
building block matrix elements
involving the $SU(3)$ color generators $T^a$:
\bea
\langle 0|\,\overline{s} \gamma^\mu T^a\left(1 - \gamma_5\right) u\,|K^+(k)
\rangle \ .
\eea
The use of quark propagators in gluon field background and
the relation
\beq
 g_s^2  G^a_{\mu\nu}G^a_{\alpha\beta} =
\frac{\pi^2}{3}\langle \frac{\alpha_s}{\pi}GG \rangle
\left(\delta_{\mu\alpha}\delta_{\nu\beta} -
\delta_{\mu\beta}\delta_{\nu\alpha}\right)
\label{gluonaverage}
\eeq
make it possible to express the non-factorizable gluonic contributions
in terms of the gluonic vacuum condensate. 
The model thus parametrizes all current and density
amplitudes in terms of \qq, \GG and $M$. The latter 
is interpreted as the constituent quark mass in mesonic fields.

In order to restrict the values of the input parameters $M$,
\qq and \GG we have studied the $\Delta I = 1/2$ selection rule
for non-leptonic kaon decay within the $\chi$QM, at $O(p^4)$
in the chiral expansion~\cite{V}.

The requirement of stability of the 
calculated isospin 0 and 2 amplitudes, togheter with
minimal $\gamma_5$-scheme dependence, selects the following range for $M$
\beq
M =  200 \pm 20 \; \mbox{MeV} \, ,
\label{M-range}
\eeq 
which is in good agreement with the range found by fitting radiative kaon 
decays~\cite{Bijnens}.

For $\Lambda^{(4)}_{QCD}$ in the range $300\div 380$ MeV
the $O(p^4)$ fit of $A_{0,2}(K^0\to\pi\pi)$
to their experimental values gives the ranges~\cite{V}
\beq
 \langle \alpha_s GG/ \pi \rangle =  
\left( 334 \pm 4  \:\: \mbox{MeV} \right) ^4 \ , 
\label{GG-range}
\eeq
\beq
\langle\bar q q \rangle = - \left(240\ ^{+30}_{-10}\:\: \mbox{MeV}\right) ^3
\label{qq-range} 
\eeq
and
\beq
M = 200\ ^{+5}_{-3}\; {\rm MeV} \ ,
\label{M-smallrange}
\eeq
for which the selection rule is reproduced with a 20\% accuracy.

\section{The $\Delta S=1$ Chiral Lagrangian}

One can write the $O(p^2)$ $\Delta S=1$
chiral lagrangian in a form which makes
the relation with the effective quark lagrangian of \eq{Lquark}
more transparent~\cite{I}:
\bea
{\cal L}^{(2)}_{\Delta S = 1}  = && 
G^{(0)}(Q_{7,8})
\Tr \left( \lambda^3_2 \Sigma^{\dag} \lambda^1_1 \Sigma
\right) \nnu \\
&+& G^{(m)} (Q_{7,8})\ \left[ \Tr \left( \lambda^3_2 \Sigma^\dag \lambda _1^1 
\Sigma {\cal M}^\dag \Sigma \right) 
 + \Tr \left(\lambda_1^1 \Sigma  \lambda^3_2  \Sigma^\dag 
{\cal M} \Sigma^\dag \right)\right] \nnu \\
& +& G_{\underline{8}} (Q_{3-10}) \Tr \left( \lambda^3_2 D_\mu \Sigma^{\dag}
D^\mu \Sigma
\right)   \nnu \\
& +&  G_{LL}^a (Q_{1,2,9,10}) \, 
\Tr \left(  \lambda^3_1 \Sigma^{\dag} D_\mu \Sigma \right)
\Tr \left( \lambda^1_2 \Sigma^{\dag} D^\mu  \Sigma \right)  \nnu \\   
& +&  G_{LL}^b (Q_{1,2,9,10})\, \Tr \left( \lambda^3_2 \Sigma^{\dag} D_\mu
\Sigma \right)
\Tr \left(  \lambda^1_1 \Sigma^{\dag} D^\mu \Sigma \right) \nnu \\
& +&  G_{LR}^a (Q_{7,8})\, \Tr \left( \lambda^3_2 D_\mu \Sigma 
\lambda^1_1 D^\mu \Sigma^{\dag} \right)  \nnu \\
& +&  G_{LR}^b (Q_{7,8})\, \Tr \left( \lambda^3_2 \Sigma^{\dag} D_\mu
\Sigma \right)
\Tr \left(  \lambda^1_1 \Sigma D^\mu \Sigma^{\dag} \right)  \nnu \\
& +&  G_{LR}^c (Q_{7,8}) \left[ \Tr \left( \lambda^3_1 \Sigma \right)
\Tr \left( \lambda^1_2  D_\mu
\Sigma^{\dag} D^\mu \Sigma\ \Sigma^{\dag} \right) \right. \nnu \\
&  &  \quad\quad\quad\quad + \left.
\Tr \left( \lambda^3_1  D_\mu \Sigma D^\mu \Sigma^{\dag}\ \Sigma\right)
\Tr \left(  \lambda^1_2 \Sigma^{\dag} \right)  \right]
\label{Lchi} \, ,
\eea
where $\lambda^i_j$ are combinations of Gell-Mann
$SU(3)$ matrices defined by $(\lambda^i_j)_{lk} = \delta_{il}\delta_{jk}$
 and ${\cal M} =$ diag($m_u,m_d,m_s$). 
The matrix $\Sigma$ is defined as
$\exp \left( \frac{2i}{f} \,\Pi (x)  \right)$
and
$\Pi (x) = \sum_a \lambda^a \pi^a (x) /2 $,
$(a=1,...,8)$ contains the pseudoscalar octet fields.
The coupling
$f$ is identified at LO with the octet decay constant.
The covariant
derivatives in \eq{Lchi} are taken with respect to the external
gauge fields whenever they are present.
Other terms are possible, but 
they can be reduced to these by means of trace identities.

The chiral coefficients $G(Q_i)$ can be determined by matching 
chiral amplitudes with those obtained in the $\chi$QM, via
integration of the quark fields at the constituent quark mass scale 
$M$. 
Given the $O(p^2)$ chiral lagrangian we can fully compute
the $O(p^4)$ meson-loop contributions to the hadronic matrix elements
of interest.
In refs.~\cite{V,I,II} one can find a detailed
description of the approach we have adopted and the results obtained.

The bosonization of the electroweak operators $Q_{7,8}$ starts with
a momentum-inde\-pen\-dent term of order $O(p^0)$ in the chiral
expansion, $G^{(0)}$; 
this is the reason behind their importance in the determination
of the ratio \ee . While the momentum-dependent $O(p^2)$ 
corrections proportional to $G_{LR}^{a,b,c}$ have
been included in our previous numerical analysis,
the momentum-independent correction proportional to $G^{(m)}$
was not determined in ref.~\cite{I} because,
although of $O(p^2)$, it vanishes in the chiral limit $m_q \rightarrow 0$
and it was erroneously thought to give a negligible contribution.

By matching the $\chi$QM calculation with that of the chiral lagrangian,
for a convenient mesonic transition,
one finds
\bea
G^{(m)}(Q_7) & = & - 3 f^2 \vev{\bar q q}/N_c \nnu \\
G^{(m)}(Q_8) & = & - 3 f^2 \vev{\bar q q} 
\label{Gm}
\eea
In ref.~\cite{FL} it was showed that $G^{(m)}$ should be proportional
to the strong lagrangian coefficient $L_8$. 
This can be verified via a strong chiral lagrangian calculation
after inclusion of the wave-function and coupling-constant
renormalization, as discussed in the appendix of ref.~\cite{V}.

The determination in \eq{Gm} yields the following corrections to the $Q_8$
hadronic matrix elements for the 0 and 2 isospin projections
of the $K^0\to\pi\pi$ transitions:
\bea
\langle Q_8 \rangle_0^{(m)} & = & 2 \sqrt{3} \frac{\vev{\bar qq}}{f} \left(
m_s + 3 m_d + 2 m_u \right) \nnu \\
\langle Q_8 \rangle_2^{(m)} & = & \sqrt{6}  \frac{\vev{\bar qq}}{f} \left(
m_s + 3 m_d + 2 m_u \right) \; .
\eea

The one-loop chiral corrections for the $K^0 \rightarrow \pi\pi$
amplitudes induced
by this term (including wave-function renormalization) are given by
\bea
a^{00}_{7,8}(\mu)&=&  - \frac{f_\pi^2\sqrt{2}}{f^5} G^{(m)} (Q_{7,8}) 
(m_s + 3 m_d + 2 m_u )
\left( 0.759 + 0.444\ i + 0.163 \ln \mu^2 \right) \\
a^{+-}_{7,8}(\mu)&=&  - \frac{f_\pi^2\sqrt{2}}{f^5} G^{(m)} (Q_{7,8})  
(m_s + 3 m_d + 2 m_u )
\left( 3.48 + 0.240\ i + 1.33 \ln \mu^2 \right) 
\eea
where $\mu$ is in units of GeV
and the labels ($00$,$+-$) refer to the decay into two neutral or charged
pions respectively.
All the relevant chiral corrections for the other terms in the
$\Delta S=1$ chiral lagrangian may be found in 
ref.~\cite{I}. 

The contribution of the $G^{(m)}$ term to
the bosonization of the electroweak operators $Q_{7,8}$ has the opposite
sign with respect to
the leading constant term. The effect of it,
combined with the inclusion of the $G_{LR}^{c}$ term
($G_{LR}^{a,b}$ are numerically negligible),
and the $O(p^2)$ part of the $\chi$QM wavefunction renormalization, 
is to reduce by 20\% the overall contribution of the electroweak
operators. This determines a sizeable increase of the predicted
central value of \ee.

\section {Hadronic Matrix Elements}

In order to perform the present analysis, 
we have computed the matrix elements of the
relevant operator to $O(p^4)$, 
the next order in the chiral expansion. Such
a computation was deemed necessary in order to further
reduce the uncertainty in
the estimate.

The total matrix elements have the form 
\beq
\langle Q_i(\mu) \rangle _{I} \; = \; Z_\pi \sqrt{Z_K}  
\left[\langle Q_i \rangle ^{LO}_{I} + 
\langle Q_i (\mu)\rangle ^{NLO}_{I} \right]  + a_i^I(\mu)\; ,
\label{hme}
\eeq
where $Q_i$ are the operators in \eq{Q1-10},
\beq
\langle Q_i \rangle _{I} \equiv
\langle (\pi \pi)_I | Q_i | K^0 \rangle\ .
\eeq
and $Z_{\pi,K}$ are the $\chi$QM wavefunction renormalizations, whose 
expressions to $O(p^4)$ can be found in ref.~\cite{V}.
The matrix elements so computed
are expanded to $O(p^4)$, discarding higher order terms.
The functions
$a_i^I(\mu)$ represent the scale dependent meson-loop corrections,
including the mesonic wavefunction renormalization. They are defined
as the isospin projections of the $a_i^{+-}(\mu)$ and $a_i^{00}(\mu)$
corrections computed in ref.~\cite{I}, properly rescaled by factors
of $f_\pi/f$ in order to replace $f_\pi\to f$ in the NLO evaluation.
The scale dependence of the NLO part of the matrix elements 
is a consequence of 
the perturbative scale dependence of the current quark masses
which enter at this order.
 
A complete list of the hadronic matrix elements to order $O(p^2)$,
namely $\langle Q_i \rangle ^{LO}_{I}$,
can be found in ref.~\cite{I}, with the proviso of replacing
all occurences of $f_\pi$ by $f$.
The NLO $I = 0$ and $2$ contributions to the $K^0\to\pi\pi$ matrix elements  
are given by:
\bea
\langle Q_1 \rangle ^{NLO}_{0} & = & \frac{1}{3} X \left[ 
 \left(-1 + \frac{2}{N_c}\right) \beta 
  - \frac{2}{N_c} \delta_{\vev{GG}} \beta_G \right]\ , \\
\langle Q_1 \rangle ^{NLO}_{2} & = & \frac{\sqrt{2}}{3} X \left[  
 \left(1+\frac{1}{N_c}\right)\beta - \frac{\delta_{\vev{GG}}}{N_c} \beta_G
 \right]\ , \\
\langle Q_2 \rangle ^{NLO}_{0} & = & -\frac{1}{3} X \left[ 
 \left(-2 + \frac{1}{N_c}\right) \beta
 - \frac{\delta_{\vev{GG}}}{N_c}  \left(\beta_G + 3 \gamma_G\right) 
\right]\ , \\
\langle Q_2 \rangle ^{NLO}_{2} & = & \langle Q_1 \rangle ^{NLO}_{2}\ , \\
\langle Q_3 \rangle ^{NLO}_{0} & = & \frac{1}{N_c} X  \left[ \beta  
 - \delta_{\vev{GG}} \left(\beta_G - \gamma_G\right)
 \right]\ , \\ 
\langle Q_4 \rangle ^{NLO}_{0} & = &\langle Q_2 \rangle ^{NLO}_{0}  
   - \langle Q_1 \rangle ^{NLO}_{0} + \langle Q_3 \rangle ^{NLO}_{0}\ , \\
\langle Q_5 \rangle ^{NLO}_{0} & = &  \frac{2}{N_c} X \; \gamma \ ,\\
\langle Q_6 \rangle ^{NLO}_{0} &=&  2 X \left[\gamma
 + \frac{\delta_{\vev{GG}}}{N_c}\gamma_G \right] \ ,\\ 
\langle Q_{7} \rangle ^{NLO}_{0} & = 
& \frac{3}{2} e_d \langle Q_{5} \rangle ^{NLO}_0 + 
 \frac{1}{2} \left(e_u - e_d \right) \left[ X \; \beta
 - 4 \sqrt{3} \frac{\langle\bar{q} q\rangle}{N_c} \; 
\gamma_E \right]
 \ , \\
\langle Q_{7} \rangle ^{NLO}_2 & = 
& - \frac{1}{2} \left(e_u - e_d \right) \left[\sqrt{2} X \beta
 + 2 \sqrt{6}\frac{\langle \bar{q} q \rangle}{N_c}\; \gamma_E
 \right]  \ , \\
\langle Q_8 \rangle ^{NLO}_0 & = & 
 \frac{3}{2} e_d \langle Q_{6} \rangle ^{NLO}_0 + 
 \frac{1}{2} \left(e_u - e_d \right)\frac{X}{N_c} 
\left[  \beta  
  + \delta_{\vev{GG}} \left( \beta_G + 3 \gamma_G \right) \right]
\nnu \\ & &
 -\  2 \sqrt{3} \left(e_u - e_d \right) 
\langle \bar{q} q \rangle \gamma_E 
\ , \\
\langle Q_8 \rangle ^{NLO}_2 & = &
 - \frac{1}{2} \left(e_u - e_d \right) \frac{\sqrt{2} X}{N_c} \left[ \beta 
 +  \delta_{\vev{GG}} \beta_G \right]
\nnu \\ & & 
 -\ \sqrt{6} \ \left(e_u - e_d \right)
\langle \bar{q} q \rangle  \gamma_E
\ , \\ 
\langle Q_9 \rangle ^{NLO}_0 & = &  
\frac{3}{2} e_d \langle Q_3 \rangle ^{NLO}_0
+ \frac{3}{2} \left( e_u - e_d \right) \langle Q_1 \rangle ^{NLO}_0\ ,\\
\langle Q_9 \rangle ^{NLO}_2 & = & 
\frac{3}{2} \left( e_u - e_d \right) \langle Q_1 \rangle ^{NLO}_2\ ,\\ 
\langle Q_{10} \rangle ^{NLO}_0 & = &  
\frac{3}{2} e_d \langle Q_4 \rangle ^{NLO}_0
+ \frac{3}{2} \left( e_u - e_d \right) \langle Q_2 \rangle ^{NLO}_0\ ,\\
\langle Q_{10} \rangle ^{NLO}_2 & = &   
\frac{3}{2} \left( e_u - e_d \right) \langle Q_2 \rangle ^{NLO}_2\ ,\\
\langle Q_{11} \rangle ^{NLO}_0 & = &
\frac{11}{4}\ X\ \frac{m_\pi^2}{m_K^2}\ \gamma_G \ , 
\eea
where $e_u=2/3$, $e_d=-1/3$ are the up-, down-quark charges respectively,
\beq
X \equiv \sqrt{3} f \left( m_K^2 - m_\pi^2 \right)
\eeq
and the gluon-condensate correction 
$\delta_{\vev{GG}}$ is given by
\beq
\delta_{\vev{GG}} =  \frac{N_c}{2} \frac{\langle
 \alpha_s G G/\pi \rangle}{16 \pi^2 f^4} \, .
\eeq

The quantities $\beta, \beta_G, \gamma$, $\gamma_G$ and
$\gamma_E$ are dimensionless 
functions of the mass parameters:
\bea
\beta & = &
 \frac{m_K^2 + 2 m_{\pi}^2}{\Lambda^2_\chi} 
-3 \frac{M}{\Lambda^2_\chi} \left(m_s + 3 \widehat m \right)   
\nnu \\
&& + \frac{m_s - \widehat m}{M} \left(1 - 6 \frac{M^2}{\Lambda^2_\chi}
      \right)  \frac{m_\pi^2}{2\left(m_K^2 - m_\pi^2 \right)}
   + \frac{\widehat m}{M}
   \left( 1 -12 \frac{M^2}{\Lambda^2_\chi}\right) \; , \\
\beta_G & = &
 \frac{m_K^2 + 2 m_\pi^2}{6 M^2} 
- \frac{5 m_s + 17\widehat m}{12 M}
- \frac{\left(m_s - \widehat m \right) m_\pi^2}{12 M 
\left(m_K^2 - m_\pi^2\right)} -\frac{\widehat m}{M}  \; , \\
\gamma & = & 
\frac{\langle \bar{q} q \rangle}{f^2} \left[
\frac{m_K^2}{ 2 M \Lambda^2_\chi} 
 - \frac{m_K^2 (2 m_s + 6 \widehat m) - m_\pi^2 (m_s + 7 \widehat m)}
{(m_K^2 - m_\pi^2) \Lambda^2_\chi}  \right. \nonumber \\
&& + \left. \frac{2 \widehat m (m_s - \widehat m )}{M (m_K^2 - m_\pi^2)}
\left( f_+ - 6 \frac{M^2}{\Lambda_\chi^2} \right) \right] \nnu \\
&& + f_+^2 \left[ - \frac{m_K^2 + m_\pi^2}{ 4 M^2 } 
 +\ \frac{( m_s + 5 \widehat m)}{2 M}  \; - \, 
\frac{2 \widehat m (m_s - \widehat m )}{m_K^2 - m_\pi^2}   \right]
\nonumber \\
&& + \frac{3 M f_+}{\Lambda^2_\chi} \left[ \frac{m_K^2}{2 M} 
 -  \;  \frac{m_K^2 (m_s + 3 \widehat m) - 2 m_\pi^2 (m_s + \widehat m)}
{(m_K^2 - m_\pi^2)}  \right] \; , \\
\gamma_G & = &
\frac{ m_K^2}{m_K^2-m_\pi^2} \frac{m_s-\widehat m}{6 M} \, . \\
\gamma_E & = & 
 \frac{\left( m_K^2 + m_\pi^2 \right)^2 + 3 m_\pi^4}{4 f M \Lambda^2_\chi} - 
 m_s \frac{\left( m_K^2 + m_\pi^2 \right) }{f \Lambda^2_\chi} - 
  \widehat m \frac{\left(2 m_K^2 + 13  m_\pi^2 \right) }{f \Lambda^2_\chi} 
\nnu \\ &&
- \frac{1}{fM}\left[ \left( m_s^2 + m_s \widehat m +
 2 \widehat m^2 \right) f_+ 
-6 \frac{M^2}{\Lambda_\chi^2}\left(m_s^2 + 2 m_s \widehat m
 + 5 \widehat m^2 \right) \right] \nnu \\ 
& &
- \frac{ff_+}{\vev{\bar{q}q}} \left(\frac{m_\pi^2}{2M} 
- 2 \widehat m \right) \left[
 f_+ \left(\frac{m_\pi^2}{2M} - m_s - 3 \widehat m \right)
+ \frac{3Mm_K^2}{\Lambda_\chi^2} \right]
 \; .
\eea

Concerning the gluonic operator $Q_{11}$, we recall that, 
due to its chiral trasformation properties, it contributes
only to the isospin zero amplitude and that
$\langle Q_{11} \rangle ^{LO} = 0$. The NLO matrix element
of $Q_{11}$ is small compared to $\langle Q_{6} \rangle ^{NLO}$
due to the additional suppression factor $m_\pi^2/m_K^2$~\cite{BEF}.

We notice that strictly speaking at $O (p^4)$ also the meson
two-loop corrections to the leading $O (p^0)$ term of the operators
$Q_{7,8}$ should be included. However, these are chirally suppressed by  
$O (M^2/\Lambda_{\chi}^{2})$ in comparison to the $O (p^4)$
contributions computed in the $\chi$QM and will be neglected.

\subsection{The $B_i$ Factors}

The effective factors
\beq
B_i^{(0,2)} \equiv \frac{\Re\left[\langle Q_i 
\rangle_{0,2}^{\chi {\rm QM}}\right]}
{\langle Q_i \rangle _{0,2}^{\rm VSA}} \, ,
\eeq
give the ratios between our hadronic matrix elements and those of the
VSA. They are a useful way of comparing different evaluations.
In Table 1, we collect the $B_i$ factors for the
relevant operators. Their values depend on the
scale at which the matrix elements
 are evaluated an on the values of the $\chi$QM input parameters; 
moreover, they depend on the
$\gamma_5$-scheme employed\footnote{The $\gamma_5$-scheme dependence enters 
at $O(p^2)$ and has been discussed in ref.~\cite{III}.}.
We have given in Table 1 a representative 
example of their values and variations in the t' Hooft-Veltman (HV)
scheme for $\gamma_5$---that we employ throughout the analysis.
\TABLE{
\begin{tabular}{|c||c|c|c|}
\hline
 & {\rm LO} & {+ $\chi$-loops} & {\rm + NLO} \\ 
\hline
$B^{(0)}_1$  & 4.2 & 7.8 & 9.5\\
\hline
$B^{(0)}_2$  & 1.3 & 2.4  & 2.9\\
\hline
$B^{(2)}_1 =B^{(2)}_2 $  & 0.60 & 0.58 & 0.41\\
\hline
$B_3$ & $-0.62$ & $-1.9$ & $-2.3$\\
\hline
$B_4$ & 1.0& 1.6 & 1.9\\
\hline
$B_5 \simeq B_6$& $1.2 \div 0.72$ & $2.0 \div 1.2 $ & $1.9 \div 1.2$\\
\hline
$B_7^{(0)} \simeq B_8^{(0)}$ &$0.91 \div 0.81$ &  $2.6 \div 2.4$ & $2.6 \div 2.4$ \\
\hline
$B_9^{(0)}$ &1.8 & 3.0 & 3.6 \\
\hline
$B_{10}^{(0)}$ &1.8 & 3.7 & 4.4\\
\hline
$B_7^{(2)} \simeq B_8^{(2)} $ &$0.81 \div 0.82$ & $0.95 \div 0.87$ & $0.94 \div 0.91$ \\
\hline
$B_9^{(2)}=B_{10}^{(2)}$ &0.60 & 0.58 & 0.41 \\
\hline
\end{tabular}
\caption{The $B_i$ factors in the $\chi$QM, leading order (LO),
 including meson-loop
renormalizations ($\chi$-loops)  
and $O(p^4)$ terms (NLO). 
We have taken the gluon
condensate at the central value of \eq{GG-range}, while the range given
for $B_{5-8}$
corresponds to varying the quark condensate according to \eq{qq-range}. 
The results shown are given in the HV scheme  for $M = 200$ MeV
and $f=86$ MeV, except for the first column where $f=f_\pi$ has been
used. 
}}

In Table 1 we have indicated by ``LO'' the complete $O(p^2)$ results,
for $f=f_\pi$,
including the $\chi$QM gluonic corrections and 
the mass term proportional to $G^{(m)}$ (discussed in section 2), 
by ``+ $\chi$-loops'' the sum of LO and chiral loop contributions,
taking $f$ at its renormalized value of 86 MeV (for a discussion
on its determination and renormalization scheme dependence 
see ref.~\cite{V}),
and by ``+ NLO'' the complete result
which includes the effect of the $O(p^4)$ $\chi$QM corrections
to the matrix elements here computed.

The large values of $B_{1,2}^{(0)}$ compared to $B_{1,2}^{(2)}$
reflect the $\Delta I = 1/2$ rule in $K^0\to\pi\pi$ decays
which is well reproduced in our approach~\cite{V}. 
The evaluation of the penguin matrix elements $\vev{Q_3}$ 
and $\vev{Q_4}$ in the $\chi$QM leads to rather large $B_i$ factors. 
In the case of $Q_3$, the $\chi$QM result has the opposite sign of the VSA
result and $B_3$ is negative.
This is the effect of the large non-factorizable gluonic
corrections. 

The
linear dependence in \qq of the gluon penguin operators $Q_6$ and $Q_5$
in the $\chi$QM matrix elements
(to be contrasted to the quadratic dependence of the VSA)  is 
responsible for the sensitivity of $B_{5,6}$ to the value
of the quark condensate.

The LO values for the electroweak penguin operators 
$Q_{7,8}$ are smaller than one because of 
the suppression induced by the $O(p^2)$ corrections 
to the leading $O(p^0)$ term discussed in section 2.

It is interesting to compare these
results with those of  lattice QCD.
The lattice estimate at the scale $\mu = 2$ GeV 
for these operators 
gives $B_5 = B_6 = 1.0 \pm 0.2$~\cite{martinelli},
$B_{7}^{(2)} = 0.58 \pm 0.02$, $B_{8}^{(2)} = 0.81 \pm 0.03$~\cite{sharpe}
  and $B_9^{(2)} = 0.62 \pm 0.10$~\cite{martinelli}. Even though the
scales are different, these factors have been shown to depend very little on
the energy scale~\cite{BJL}. We can therefore notice
the qualitative agreement, at least for $B_{7,8,9}$, on the fact that all
values are smaller than the corresponding VSA.
A lattice estimate of $B_3$ would be a desirable check of the modeling 
of non-factorizable contributions in the $\chi$QM.

\section{The Mixing Parameter $\mbox{Im}\, \lambda_t$}

In this section we update the
determination of the combination of KM entries 
$\Im V^*_{ts}V_{td}$ which is crucial in the study
of \ee. Such an update is required in the light of
our recent $O(p^4)$
estimate of the parameter $\widehat B_K$ given in ref.~\cite{V}.

A range for Im $\lambda_t$ is determined from the experimental value of
$\varepsilon$ as a function of $m_t$ and the other
relevant parameters involved in the theoretical estimate.
We will use the recent NLO results for the QCD correction factors
$\eta_{1,2,3}$~\cite{NLO2} 
and vary the $\Delta S = 2$ hadronic parameter $\widehat B_K$ 
around the central value obtained within the $\chi$QM~\cite{V}.

In order to restrict the allowed values of Im $\lambda_t$ we  solve
 the equations
\beq
\varepsilon^{\rm th} 
 ( \eta,\rho, |V_{cb}|, |V_{us}|, 
   \Lambda_{\rm QCD}, m_t, m_c, \widehat{B}_K ) =  
 \varepsilon   
\label{bound1}
\eeq
and
\bea
\eta^2 + \rho^2 &=&  \frac{1}{| V_{us}|^2} \frac{|V_{ub}|^2}{|V_{cb}|^2}
\label{bound2} \\
\eta^2 \left(1- \frac{|V_{us}|^2}{2}\right)^2 + 
\left[1-\rho \left(1- \frac{|V_{us}|^2}{2}\right) \right]^2 &=&  
\frac{1}{| V_{us}|^2} \frac{|V_{td}|^2}{|V_{cb}|^2}\ ,
\label{bound3} 
\eea
in terms of the $\eta$ and $\rho$ parameters of the Wolfenstein
parametrization of the KM matrix.

We thus
find the allowed region in the $\eta-\rho$ plane, given $m_t$, $m_c$
 and~\cite{PDG}  
 \bea
 |\varepsilon| & = & (2.266 \pm 0.023) \times 10^{-3}  \\
 |V_{us}| & = & 0.2205 \pm 0.0018  \\
 |V_{cb}| & = & 0.040 \pm 0.003  \\
 |V_{ub}|/|V_{cb}| & = & 0.08 \pm 0.02 \, . 
\eea
For $|V_{td}|$ we use the
bounds provided by the measured $\bar B_d^0$-$B_d^0$
mixing according to the relation~\cite{munich}
\beq
|V_{td}| = 8.8\cdot 10^{-3} 
\left[\frac{200\ {\rm MeV}}{\sqrt{B_{B_d}} F_{B_d}}\right]
\left[\frac{170\ {\rm GeV}}{m_t(m_t)}\right]^{0.76}
\left[\frac{\Delta M_{B_d}}{0.50/{\rm ps}}\right]^{0.5}
\sqrt{\frac{0.55}{\eta_B}}\ ,
\label{Vtd}
\eeq
with the ranges for the various parameters given in appendix A.

The theoretical uncertainty on the hadronic  
$\bar K^0-K^0$ matrix element 
controls a large part of the uncertainty on the determination
of $\Im \lambda_t$.
 For the renormalization group invariant parameter
 $\widehat B_K$  we take the range found  to $O(p^4)$
in the $\chi$QM~\cite{V}:
 \beq
 \widehat B_K = 1.1 \pm 0.2 \, ,
 \label{BKrange} 
 \eeq
where the error of the above direct determination
includes also the whole range of values extracted independently
from the study of the $K_L^0$-$K_S^0$ mass difference~\cite{V,IV}.  

The NLO order QCD corrections $\eta_{1,2,3}$ in the $\Delta S=2$
Wilson coefficient are computed by taking
$\Lambda^{(4)}_{\rm QCD} = 340 \pm 40$ MeV~\cite{PDG}, 
$m_b(m_b)=4.4$, $m_c(m_c) =1.4$ and
 $m_t^{\rm (pole)} = 175 \pm 6$ GeV~\cite{mt}, which in LO corresponds
to $m_t(m_W) = 177 \pm 7$ GeV, where the running masses
are given in the $\overline{MS}$ scheme. 
As an example, for central values of the input parameters
we find at $\mu=m_c$
\beq
\eta_1 = 1.33\ ,\quad \eta_2 = 0.51\ ,\quad \eta_3= 0.44 \, .
\eeq

\FIGURE{              
\epsfxsize=8cm
\centerline{\epsfbox{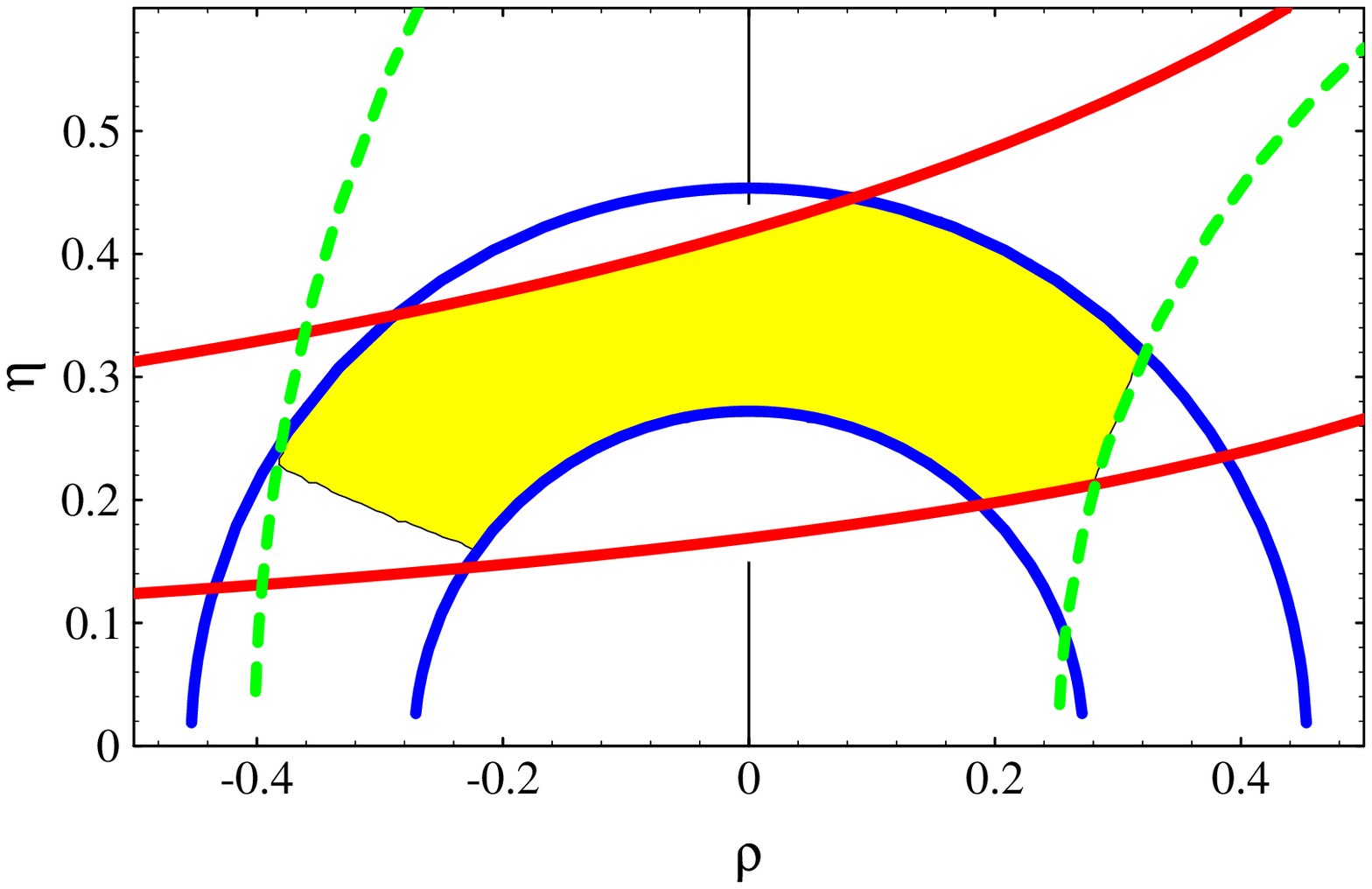}}
\caption{Range of allowed ($\eta$,$\rho$) values for 
$m_t^{\rm (pole)}$ = 169 GeV. See the text for explanations.}}
\FIGURE{              
\epsfxsize=8cm
\centerline{\epsfbox{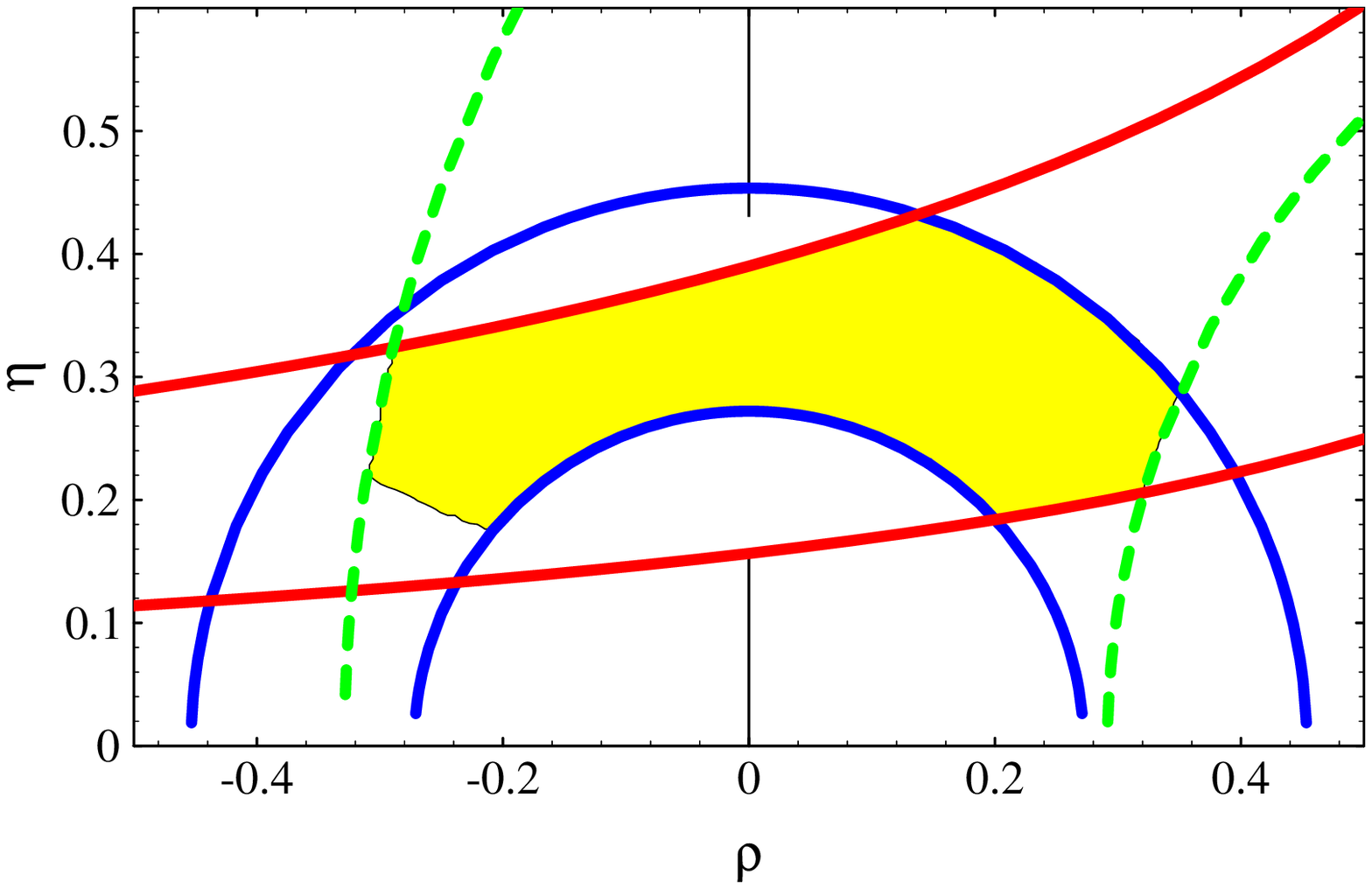}}
\caption{Same as in Fig.\ 1 for $m_t^{\rm (pole)}$ = 181 GeV.}}
 
Figs.\ 1 and 2  give the results of our analysis for the
two extreme values of $m_t$: the area enclosed
by the two black 
circumferences represents the constraint of \eq{bound2}, 
the area between the two gray (dashed)
circumferences is allowed by the bounds from
\eq{bound3}; the area enclosed by the two solid parabolic curves 
represents the solution of \eq{bound1} with
the range of $\widehat B_K$ given in \eq{BKrange} (notice that the upper
parabolic curve corresponds to the minimal value of $V_{cb}$ and
vice versa for the lower curve).

The gray region within the intersection of the curves is the range actually
allowed after the correlation in $V_{cb}$ between \eq{bound1} and
\eq{bound3} is taken into account. A further correlation is present 
when computing Im $\lambda_t$ from $\eta$ in \eq{imt}. 

This procedure determines the allowed range
for 
\beq
\Im \lambda_t \simeq \eta |V_{us}| |V_{cb}|^2 \, . 
\label{imt} 
\eeq
For a flat variation of the input parameters, including
$\Lambda^{(4)}_{\rm QCD}$, we find
 \beq
 0.62 \times 10^{-4} \leq \Im \lambda_t \leq 1.4 \times 10^{-4}\ .
\label{imlamti1} 
\eeq 
For comparison,
values of Im $\lambda_t$ between $0.86 \times 10^{-4}$ and
 $1.7 \times 10^{-4}$ are found in the most recent update of
the Munich group~\cite{munich} using  
$\widehat B_K = 0.75 \pm 0.15$.
The larger values of $\widehat B_K$ we obtain reduce substantially
the maximum values allowed for Im $\lambda_t$.

\clearpage
\section{Estimating \ee}

\FIGURE{
\epsfxsize=8cm
\centerline{\epsfbox{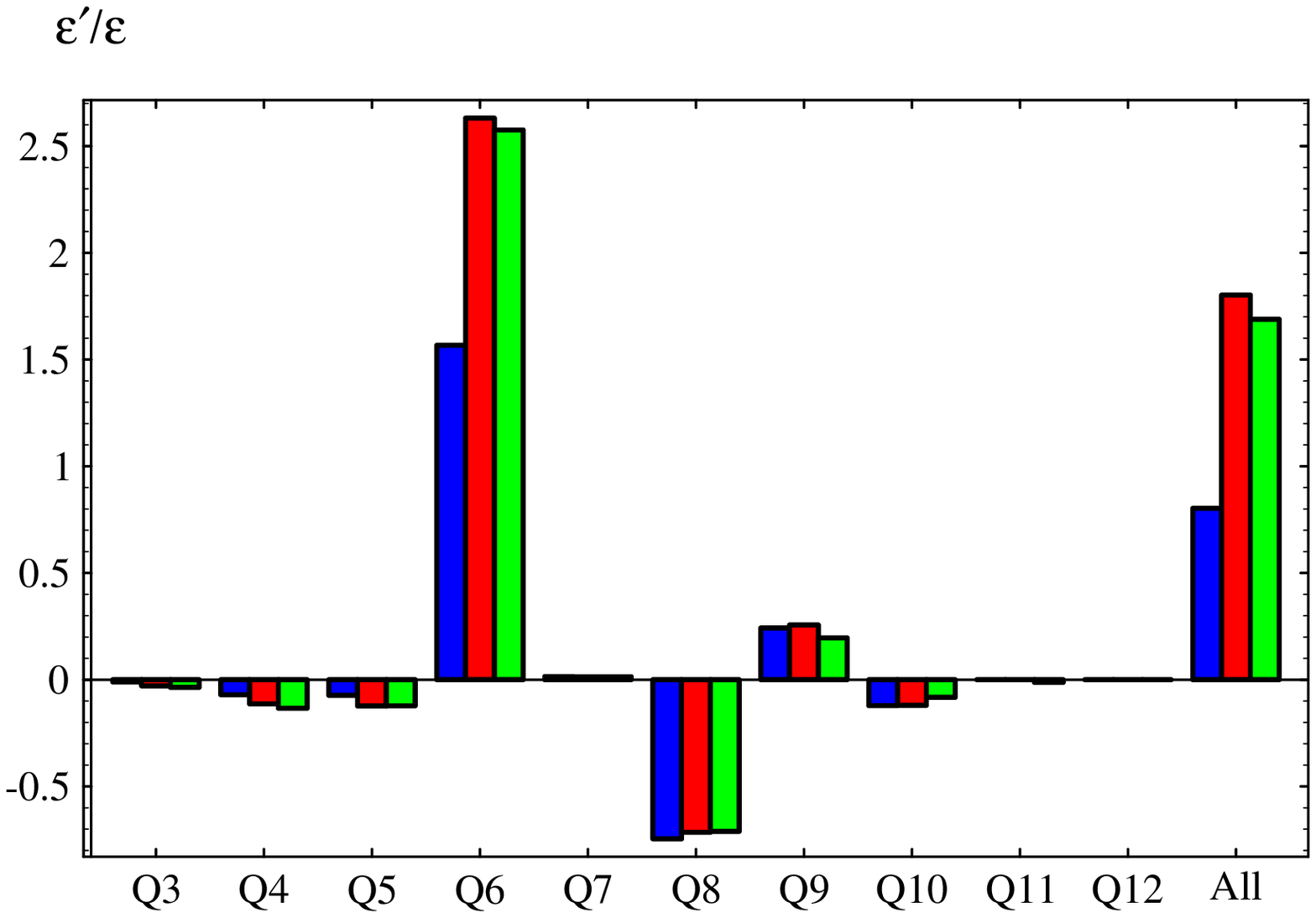}}
\caption{Anatomy of \ee in units of $10^{-3}$ 
for central values of the input parameters:
LO calculation (black), LO with chiral loops (half-tone),
complete NLO result (gray). 
In the LO case we have taken $f=f_\pi$,
whereas in the remaining hystograms the renormalized value $f=86$ MeV has
been used.}
}

We can now discuss our results for \ee.
Fig.\ 3  shows the impact 
on the final value of \ee of each operator for the representative
central values of all input parameters and 
$\Im \lambda_t = 1.0 \times 10^{-4}$. The same
figure also shows how the results vary from
the LO predictions (black), once chiral loops (half-tone) and
 NLO  (gray) corrections are included. 
The typical size of the $\chi$QM 
NLO corrections to the matrix elements of the quark
operators is less than 10\%. 

The contribution of the chromomagnetic penguin $Q_{11}$ is very small
for two reasons, as discussed in ref. \cite{BEF}:
its matrix element is vanishing at the LO, and the
NLO result turns out to be proportional to $m_\pi^2$. 
Numerically, the matrix element of $Q_{11}$ is about 5\% of the NLO 
corrections to the matrix element of $Q_6$.

The LO reduction of the contribution of the operator $Q_8$ caused by the
$O(p^2)$ correction terms
is an important result of the present analysis. In fact, without them the
contributions of $Q_6$ and $Q_8$ are at LO almost equal in absolute value
thus leading to a vanishing value for \ee. 
With the inclusion of the $O(p^2)$ corrections
the final value of \ee turns out to be positive
in the whole parameter range.
It is important to remark that this result is also
determined by the enhancement of the $Q_6$
matrix element due to the chiral loop corrections.
In turn, this effect is  
related to the good fit of the amplitude $A_0(K^0\to\pi\pi)$
which we obtain in the same framework~\cite{V}.

\FIGURE{              
\epsfxsize=8cm
\centerline{\epsfbox{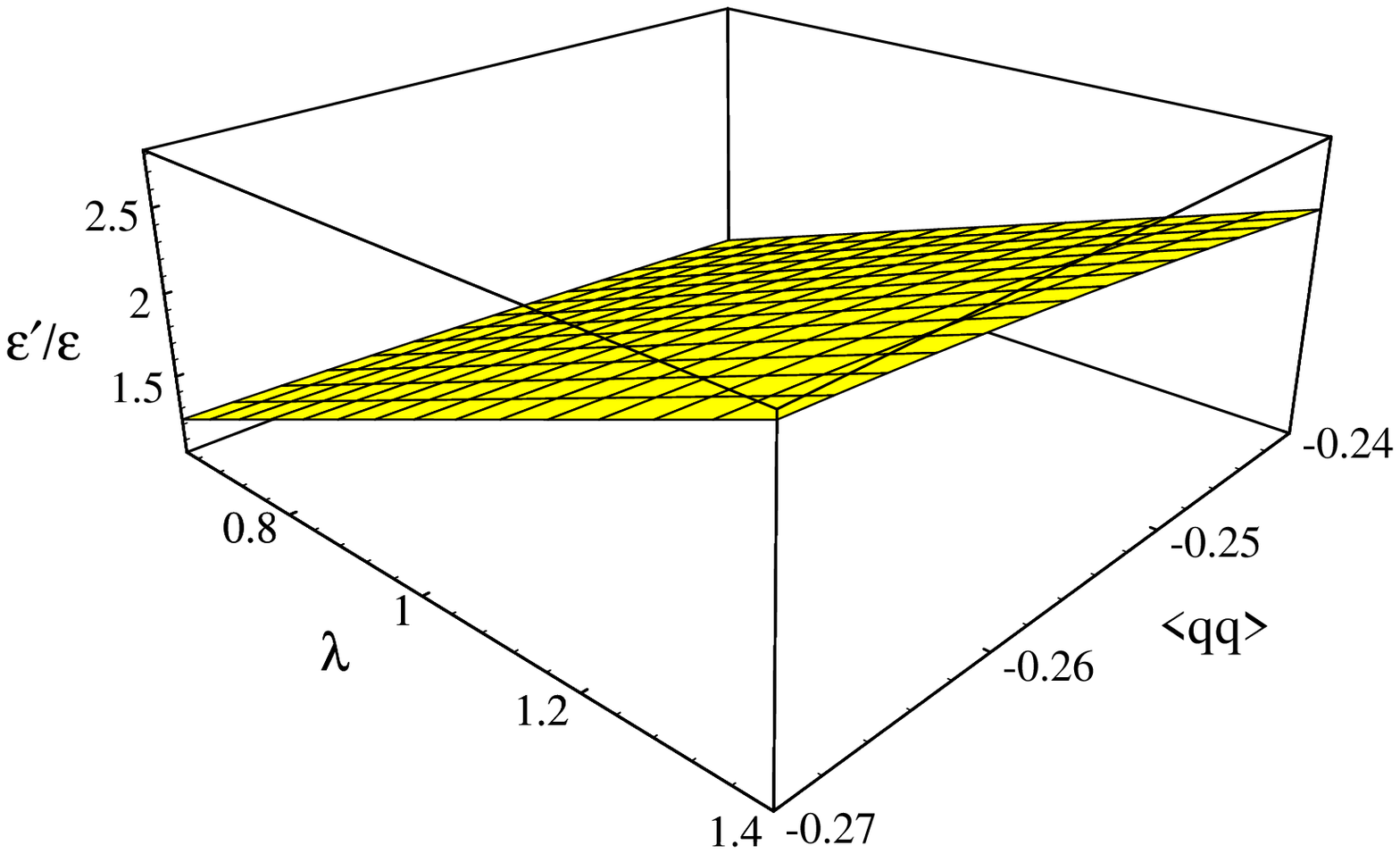}}
\caption{\ee in units of $10^{-3}$ as a function of \qq$^{1/3}$ (GeV) 
and $\lambda=$ Im $\lambda_t$ in units of $10^{-4}$,
for central values of the other input parameters. }
}

\FIGURE{              
\epsfxsize=8cm
\centerline{\epsfbox{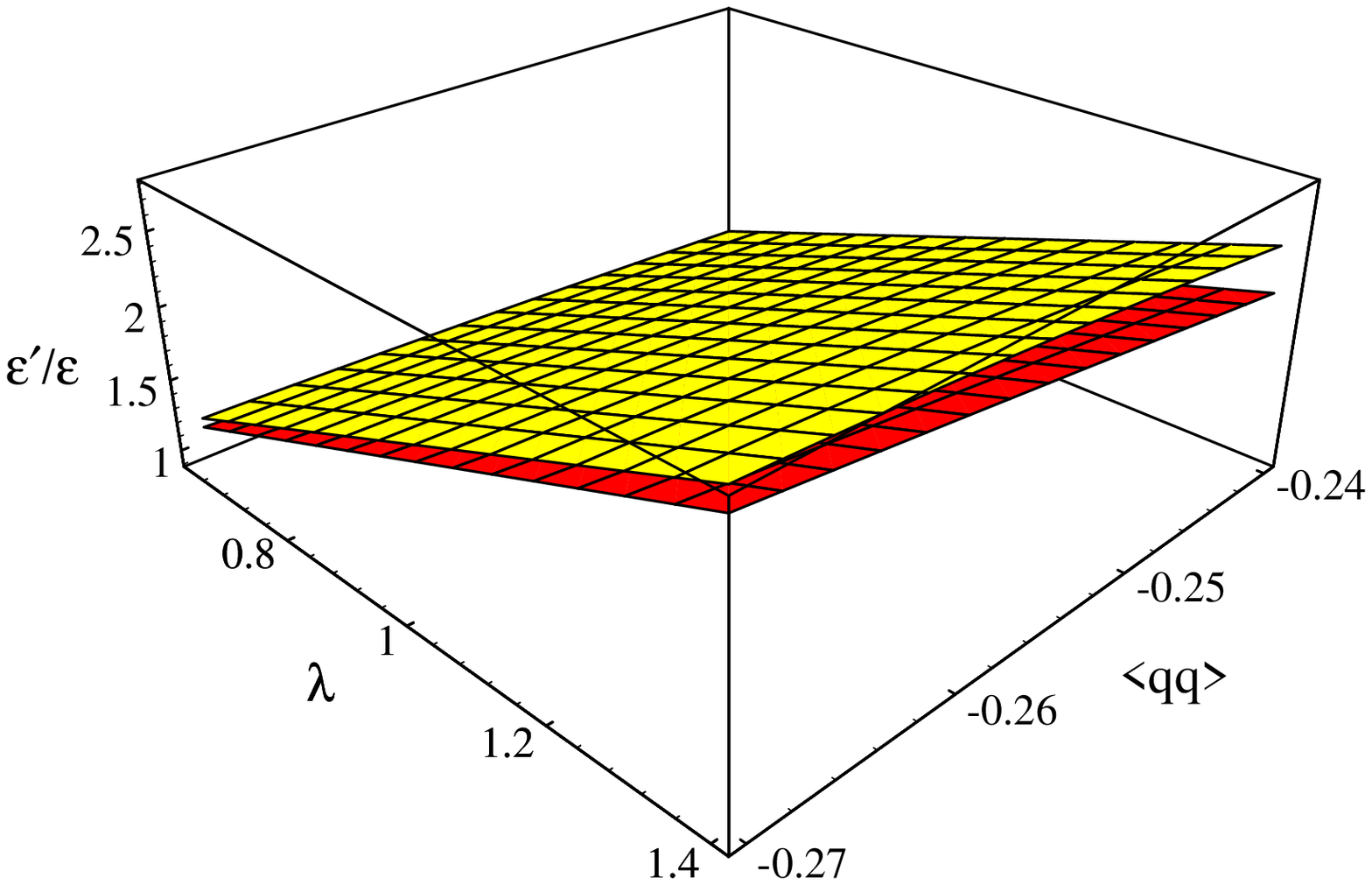}}
\caption{The dependence of \ee in units of $10^{-3}$
on the matching scale $\mu$  as a function of \qq$^{1/3}$ (GeV) 
and $\lambda=$ Im $\lambda_t$ in units of $10^{-4}$.
The gray surface corresponds to 
$\mu= 0.8$ GeV and the dark surface to $\mu= 1$ GeV. }
}

In order to understand the uncertainty in our estimate,
we now present a detailed study of the determination of \ee .
The matrix elements depend on the values of the input
parameters \qq and \GG, besides that of the constituent quark
mass $M$ characteristic of the $\chi$QM.

The range of values for \qq, \GG and $M$ found in ref.~\cite{V} 
via the NLO best fit of the $\Delta I = 1/2$ rule are reported in
\eqs{GG-range}{M-smallrange}.
Since the value of \GG is not very important in the physics
of penguin operators we can safely consider the central value of
\eq{GG-range} as a fixed input for our numerical analysis of \ee.

A relevant parameter 
in determining the size of \ee is $\Im\lambda_t$, as
discussed in the introduction. We use the overall
range given by \eq{imlamti1}.

Fig.\ 4 shows \ee  as a function of \qq and 
Im $\lambda_t$ for central values of the other input parameters.
The stability of the result can be gauged by comparing the variation
of the surface in Fig.\ 4 as we vary 
\begin{itemize}
\item The matching scale $\mu$, between 0.8 and 1 GeV: Fig.\ 5; 
\item The values of $\Lambda_{\rm QCD}$ and $m_t$ in the ranges
given in the appendix A: Fig.\ 6;
\item The value of $M$ in the range
of \eq{M-smallrange}: Fig.\ 7.
\end{itemize}

As we can see from Fig. 5,
the dependence on the matching scale
is very weak and vanishes within the range of \qq shown. 
We consider the scale stability a success of our approach.

\FIGURE{              
\epsfxsize=8cm
\centerline{\epsfbox{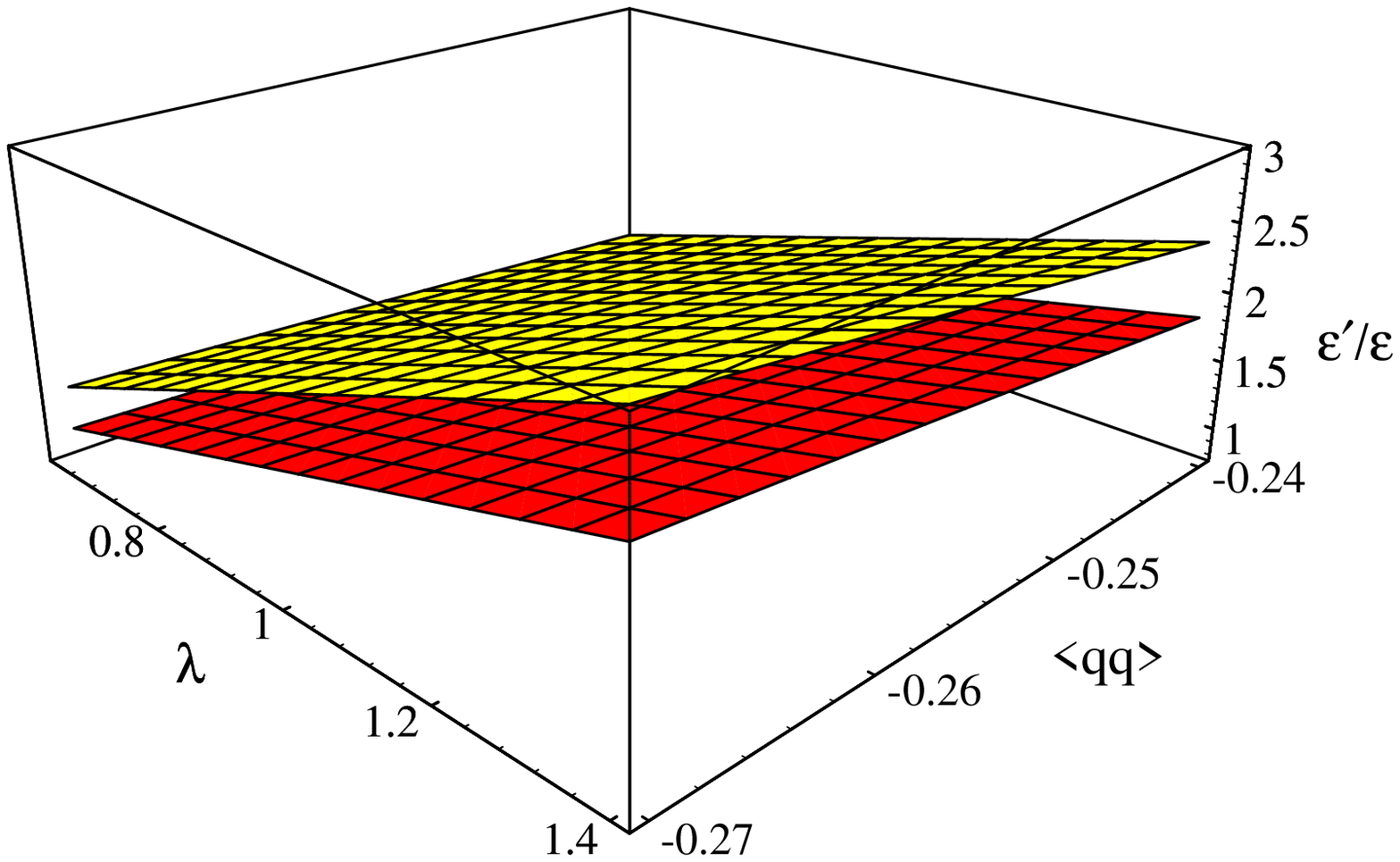}}
\caption{The dependence of \ee (in units of $10^{-3}$) 
on $\Lambda_{\rm QCD}$ and $m_t$ as a function of \qq$^{1/3}$ (GeV) 
and $\lambda=$ Im $\lambda_t$ in units of $10^{-4}$.
The upper surface corresponds to 
$\Lambda^{(4)}_{\rm QCD}=380$ MeV and $m_t^{\rm (pole)}=169$ GeV
while the lower surface corresponds to
$\Lambda^{(4)}_{\rm QCD}=300$ MeV and $m_t^{\rm (pole)}=181$ GeV.}
}

\FIGURE{              
\epsfxsize=8cm
\centerline{\epsfbox{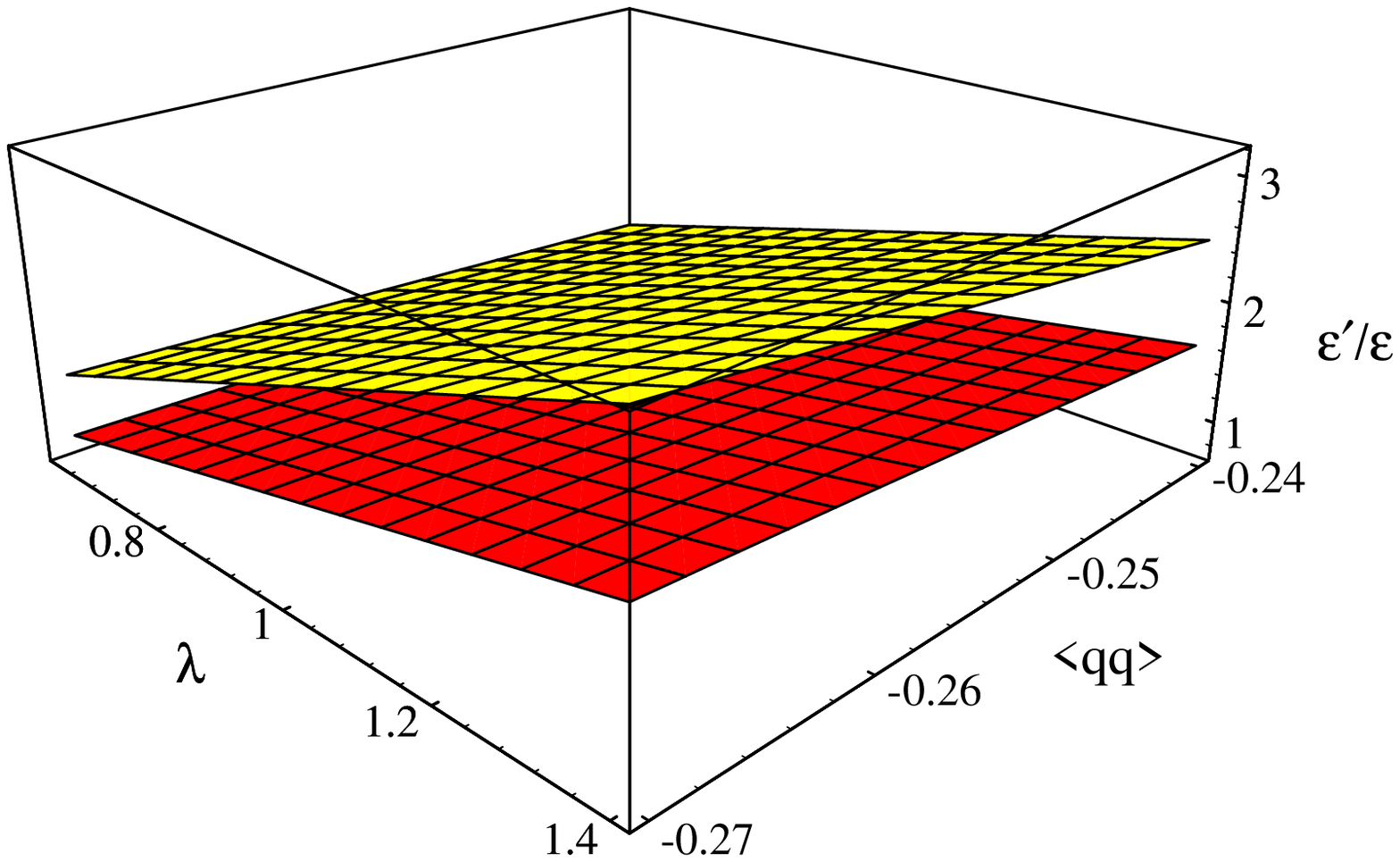}}
\caption{The dependence of \ee in units of $10^{-3}$ on the constituent
quark mass $M$ as a function of \qq$^{1/3}$ (GeV) 
and $\lambda=$ Im $\lambda_t$ in units of $10^{-4}$.
The upper surface corresponds to $M= 197$ MeV, 
$\Lambda^{(4)}_{\rm QCD}=380$ MeV and $m_t^{\rm (pole)}=169$ GeV,
while the lower surface corresponds to $M= 205$ MeV,
$\Lambda^{(4)}_{\rm QCD}=300$ MeV and $m_t^{\rm (pole)}=181$ GeV.}
}

By varying $\Lambda_{\rm QCD}$, $m_t$, \qq and
$\Im \lambda_t$ in their respective ranges while keeping $M$ fixed at its
central value, we find (Fig. 6)
\beq
\varepsilon '/\varepsilon  = 
 1.7\ ^{+1.3}_{-0.9} \,\times \,10^{-3} \, .
 \label{best} 
\eeq

The final source of
uncertainty arises from the value of $M$.
If we include its variation according to the range of \eq{M-smallrange}
we find (Fig. 7)
\beq
\varepsilon '/\varepsilon  = 
 1.7\; ^{+1.4}_{-1.0} \,\times \,10^{-3} \, .
 \label{worst} 
\eeq
The result in \eq{worst} 
represents the most conservative estimate of the error in
our prediction. 
The dependence on the remaining input parameter
$m_s$ in the NLO corrections is small and can be safely neglected.

In comparing the present estimate with our previous one 
in ref.~\cite{III} (circa 1995), 
we may notice that the complete
$O(p^2)$ analysis of the bosonization of the operator $Q_8$ has reduced the
impact of the electroweak corrections and accordingly \ee 
is never negative. In addition, the inclusion in the present
analysis of the effects
of the rescattering phases, \eqs{PI0}{PI2}, enhances the $I=0$ channel,
thus increasing the size of \ee.
As a consequence, the values of \ee are now larger
even though the NLO determination of the parameter 
$\widehat B_K=1.1\pm 0.2$ has made the maximum value of 
Im $\lambda_t$ roughly 30\% smaller. 

Finally, the 
updated new short-distance analysis,
with the reduced range in $\Lambda^{(4)}_{\rm QCD}$, 
makes the whole estimate more stable.

\subsection{Outlook}

Our analysis, 
based on the  implementation
of the $\chi$QM  and chiral lagrangian methods,
takes advantage of the observation that the $\Delta I = 1/2$ selection
rule in kaon decays is well reproduced in terms of three basic 
parameters (the constituent quark mass $M$ and the quark and gluon 
condensates) in terms of which all hadronic matrix elements
of the $\Delta S=1$ lagrangian can be expressed.
We have used the best fit of the $\Delta I=1/2$
selection rule in kaon decays to constrain the allowed
ranges of $M$, $\vev{\bar q q}$ and $\vev{GG}$ and  have fed them
in the analysis of $\varepsilon '/\varepsilon$ based on the
NLO $O(p^4)$
 determination of all hadronic matrix elements. 
Values of \ee positive and
of the order of $10^{-3}$  
are preferred, even though values of $O(10^{-4})$
cannot be excluded.

\FIGURE{              
\epsfxsize=8cm
\centerline{\epsfbox{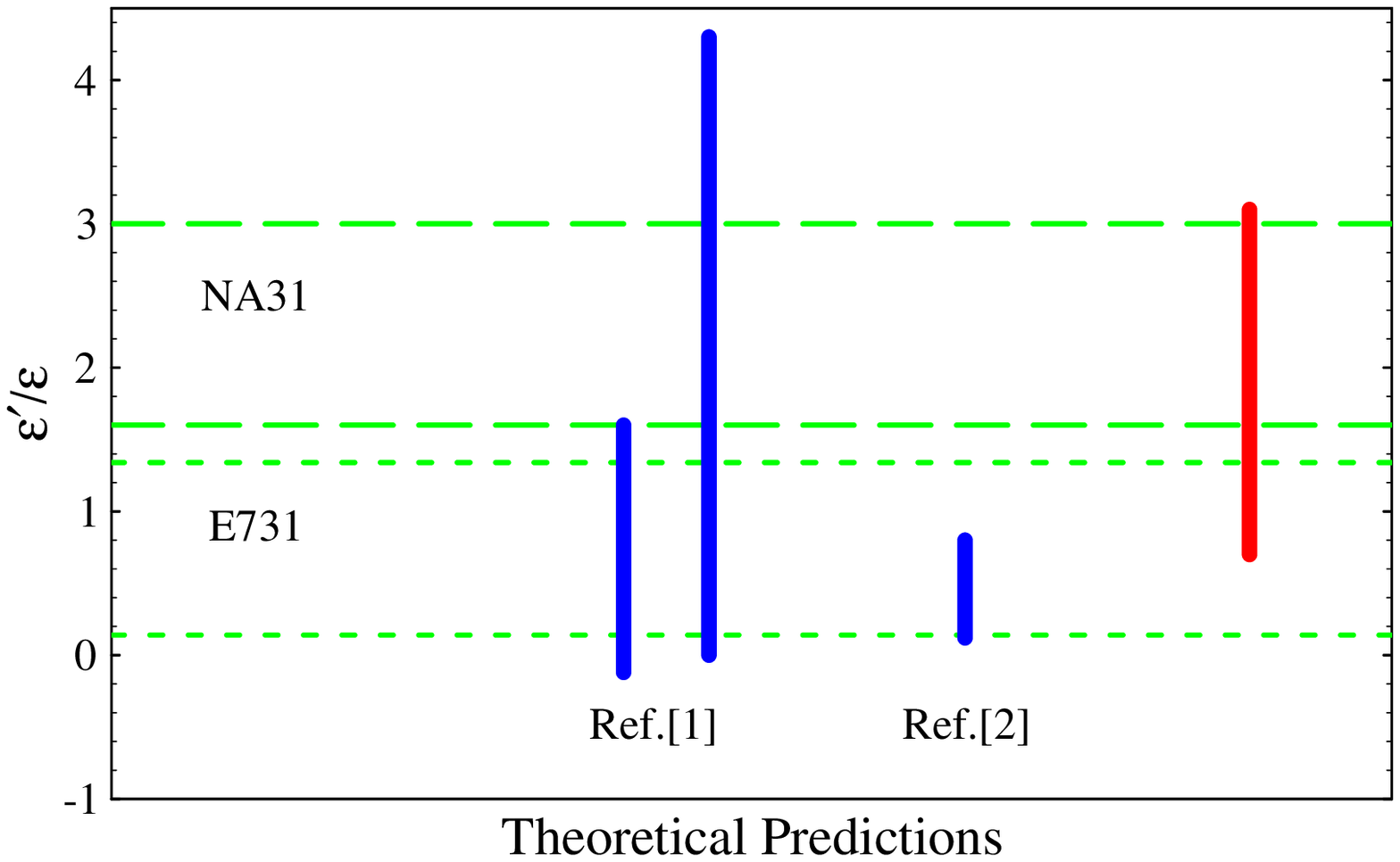}}
\caption{Present status of theoretical predictions and experimental values
for $\varepsilon'/\varepsilon$ (in units of $10^{-3}$). 
The two horizontal bands bounded by long- and short-dashed lines
show the 1$\sigma$ experimental results of NA31 and E731 respectively.
The vertical bars represent the theoretical predictions including the errors
quoted by the authors.
In gray we show our final estimate of \eq{worst}.}}

In Fig.\ 8 we have summarized the present 
status of the theoretical predictions
for $\varepsilon'/\varepsilon$. The two ranges we have reported
for ref.~\cite{munich} 
corresponds to taking two different determinations of
$m_s$, namely, from left to right, 
$m_s(1.3\ {\rm GeV})=150\pm 20$ MeV and
$m_s(1.3\ {\rm GeV})=100\pm 20$ MeV.
The varying of $m_s$ in the approach of ref.~\cite{munich}
is the analogue of varying \qq in our LO matrix elements.

The smaller uncertainty in the lattice result~\cite{rome}
is due to the authors' 
Gaussian treatment of the uncertainties 
in the input parameters, to be contrasted to a flat scanning. 
The results of ref.~\cite{munich} as well as ours
 can be made proportionally smaller by a similar
treatment of the errors.   

Given the complexity of the computation, it is rewarding to find that
so very different approaches lead to predictions that are reasonably
in agreement with each other.

We would like to conclude by briefly discussing
 three issues that arise in
 comparing our present analysis with that of refs.~[1] and [2]:
\begin{itemize}
\item 
The weight of the operator $Q_4$, that is negligible in our
estimate (Fig. 3), 
is made sizable in those of refs.~[1] and [2] by the assumption that
the coefficient $B_4$ can be as large as 6. Such a large value comes from
considering the relation
\beq
Q_4 = Q_2 - Q_1 + Q_3\ ,
\eeq
which is exact in the HV scheme,
and the large value of the difference $\vev{Q_2}_0 - \vev{Q_1}_0$ that is 
required in order to account for the $\Delta I =1/2$ rule. The $\chi$QM
result of a negative $B_3$ shows that a large 
$\vev{Q_2}_0 - \vev{Q_1}_0$ does not
necessarily implies a correspondingly large value of $B_4$.
A large value of $B_4$ makes \ee smaller 
because the operator $Q_4$ gives a contribution of the opposite sign 
with respect to that of $Q_6$.
\item 
In ref.~[1] only the terms proportional to $G^{(0)}$ and $G^{b}_{LR}$ of the
$\Delta S=1$ chiral lagrangian are included
in the matrix elements of the operators $Q_{7,8}$. While the effect of the
missing terms $G^{a,c}_{LR}$ 
is within the variation of the coefficient $B_8$ that
is there considered, the central value of \ee may be affected by such a choice.
\item 
It is difficult to compare our approach to that of the lattice.
We however notice that the contribution of the chiral term proportional to 
$G^{c}_{LR}$ only starts at the level of the $K \rightarrow 2 \pi$ amplitude
and therefore is not included when computing the two-point 
$K \rightarrow \pi$ transition, as currently done on the lattice. 
We would also
like to stress the importance of always using the full $O(p^2)$ $\Delta S=1$
chiral lagrangian~(\ref{Lchi}) in computing the $K \rightarrow 2 \pi$ 
from the $K \rightarrow \pi$ amplitude.
\item
Finally, the 20\% enhancement of the $I=0$ amplitude due to the rescattering
phase is included only in our analysis. 
\end{itemize}

\acknowledgments{Work 
partially supported by the Human Capital and Mobility EC program under 
contract no. ERBCHBGCT 94-0634. JOE thanks SISSA for its ospitality.}

\clearpage
\appendix
\section{Input Parameters} 
\begin{table}[h]
\begin{center}
\begin{footnotesize}
\begin{tabular}{|c|c|}
\hline
{\rm parameter} & {\rm value} \\
\hline
$\sin ^2 \theta_W(m_Z)$ & 0.23 \\
$m_Z$ & 91.187 GeV \\
$m_W$ & 80.33 GeV \\
$m_t^{\rm (pole)}$ & $175 \pm 6$ GeV \\
$m_t(m_t)$ & $167 \pm 6$ GeV \\
$m_t(m_W)$ & $177 \pm 7$ GeV \\
$m_b(m_b)$ & 4.4 GeV \\
$m_c(m_c)$ & 1.4 GeV \\
$m_s$ (1 GeV) & $178 \pm 18$ MeV \\
$m_u + m_d$ (1 GeV) & $12 \pm 2.5$ MeV \\
$\Lambda_{\rm QCD}^{(4)}$ & $340 \pm 40$ MeV \\
\hline
$V_{ud}$ & 0.9753 \\
$V_{us}$ & $0.2205 \pm 0.0018$ \\
$ |V_{cb}|$ & $0.040 \pm 0.003$ \\
$|V_{ub}/V_{cb}|$ & $0.08 \pm 0.02$ \\
\hline
$\eta_B$ & $0.55\pm 0.01$ \\
$\sqrt{B_{B_d}} F_{B_d}$ & $ 200\pm 40$ MeV \\
$\Delta M_{B_d}$ & $ 0464 \pm 0.018$ ps$^{-1}$ \\
\hline
$\widehat B_K$ & $1.1 \pm 0.2$ \\ 
$\Im \lambda_t$ & $ (1.0 \pm 0.4 ) \times 10^{-4} $ \\
\hline
$M$ &   $  200\ ^{+5}_{-3} \: \mbox{MeV}$ \\
$\vev{\bar{q}q}$  &  $- (240\ ^{+30}_{-10} \: \mbox{MeV} )^3$ \\
$ \langle \alpha_s GG/\pi \rangle $ & $(334 \pm  4 \: 
\mbox{MeV} )^4 $ \\
\hline
$f$  &  86 $\pm$ 13  MeV \\
$f_\pi = f_{\pi^+}$  &  92.4  MeV \\
$f_K = f_{K^+}$ & 113 MeV \\
$m_\pi = (m_{\pi^+} + m_{\pi^0})/2 $ & 138 MeV \\
$m_K = m_{K^0}$ &  498 MeV \\
$m_\eta$ & 548 MeV \\
$\Omega_{\eta+\eta'}$ & $0.25\pm 0.05$ \\
$\cos\delta_0$ & $0.8$ \\
$\cos\delta_2$ & $1.0$ \\
\hline
\end{tabular}
\end{footnotesize}
\end{center}
\caption{Table of the numerical values of the input parameters.}
\end{table}

\clearpage
%
%
\vspace{1cm}
%
%
\renewcommand{\baselinestretch}{1}

\end{document}